\documentclass[10pt, preprint]{aastex}

\usepackage{graphics}
\usepackage[hyperfootnotes=false]{hyperref}
\usepackage{color} 
\usepackage{amsmath}

\newcommand{\unit}[1]{\ensuremath{\, \mathrm{#1}}}

\def\beq{\begin{equation}}
\def\eeq{\end{equation}}

\def\ea{{\it et al.\ }}

\newcommand{\bea}{\begin{eqnarray}}
\newcommand{\eea}{\end{eqnarray}}
\newcommand{\eiso}{{\cal E}_{\rm iso}}
\newcommand{\egamma}{{\cal E}_\gamma}
\newcommand{\ep}{E_p}

\newcommand{\swift}{{\it Swift \,}}
\newcommand{\swiftA}{{\it Swift}}
\begin{document}

\title{Cosmological Evolution of Long Gamma-ray Bursts and Star Formation Rate}
\author{Vah\'e Petrosian$^{1,2}$\, Ellie  Kitanidis$^{3}$
and Daniel Kocevski$^4$
}
\affil{$^{1}$Department of Physics and KIPAC, Stanford University,
Stanford, CA 94305, USA\\ 
$^{2}$Department of Applied Physics, Stanford University, Stanford, CA 94305,
USA\\ 
$^{3}$Department of Physics, UC Berkeley, Berkeley, CA USA\\
$^4$NASA Goddard Space Flight Center, College Park MD
}
\shorttitle{Evolution of Gamma-ray  Bursts}
\shortauthors{Petrosian \&  Kitanidis}

\begin{abstract}

Gamma-ray bursts (GRBs) by virtue of their high luminosities can be
detected up to very high redshifts and therefore can be excellent probes of the
early universe. This task is hampered by the fact that most of their
characteristics have a broad range so that we first need to obtain an
accurate description of the distribution of these characteristics, and
specially, their cosmological evolution. We use a sample of about 200 \swift
long GRBs with known redshift to determine the luminosity and formation rate
evolutions and the general shape of the luminosity function. In contrast to most
other forward fitting methods of treating this problem we use the Efron
Petrosian methods which allow a non-parametric determination of above
quantities. We find a relatively strong luminosity evolution, a luminosity
function that can be fitted to a broken power law, and an unusually high rate of
formation rate at low redshifts, a rate more than one order of magnitude higher
than the star formation rate (SFR). On the other hand, our results seem to agree
with  the almost constant SFR in redshifts 1 to 3 and the decline above this
redshift.

\end{abstract}


\section{Introduction}
\label{intro}

Observations of increasing numbers of  gamma-ray bursts
({\bf GRBs})  with measured  redshifts, up to $z\sim 10$, by instruments on
board {\it
BeppoSAX}, {\it HETE}, {\it INTEGRAL}, and in particular 
{\it Swift}, have stimulated many uses of them as cosmological probes
either as ``standard Candles" (SC) for determination of global
cosmological parameters, such as density parameters  and  equations of state, or
as
probes of  the early phases of the
universe such as reionization, or star formation
rate (SFR) and cosmic metalicity evolution (CME) at high redshifts.
Unfortunately, most intrinsic {\it distance dependent} charcteristics of GRBs,
such as their peak luminosity, total emitted energy, etc, have a broad
distribution making them unsuitable as SC. However,  there has been several
attempts to discover some distance dependent characteristic  that
shows a well defined correlation
with another {\it distance independent}
characteristic, which can then be used to determine distances
as in Cepheid variables or  type Ia supernovae.
Example of such relations for GRBs are the  correlations between 
lag and luminosity (Norris et al. 2000), the variability and luminosity
(Reichart et al. 2001), and  the
peak energy $\ep$ of the $\nu F_\nu$ spectrum and the total (isotropic)
gamma-ray  energy
$\eiso$ (Amati et al.  2002; Lamb et al. 2004; Attiea et al. 2004) or  the
beaming corrected
energy  $\egamma$ (Ghirlanda et al. 2004). For a more general review see
Xiao \& Schaffer (2009).

There are, however, many uncertainties in the claimed relations rendering 
the cosmological tests unreliable. 
First, it is now clear that these characteristics have  broad distributions and
that the correlations are not  simple one-to-one relations that is sometimes
claimed and  do not have the small dispersion  required for precision
cosmological
tests. Instead, most of the correlations are 
statistical in nature, as originally predicted  by Lloyd et al. (2000)
long before any redshifts were measured, and not valid for the GRB population as
a whole (Nakar \& Piran 2004; Band \& Preece 2005). Butler et al. (2009) stress
that  a
careful
accounting of \underline{observational selection effects} is required, and
suggest using the
methods developed  by Efron and Petrosian (EP for short, see below), to quantify
the nature
and degree of the correlations before any cosmological tests can be affected.

Secondly,
even if there
exists a one-to-one  relation;  e.g.  $\eiso = {\cal E}_0 {\cal C}(
\ep/E_0)$,
the relation between the cosmological parameters, such as matter and dark
energy density parameters $\Omega_m$ and $\Omega_\Lambda$ and observed
quantities, such as redshift $z$,  flux $f(t)$ [or fluence $F=\int f(t)dt$] and
$\ep$, are complicated.  Here $\eiso$ is related to the total gamma-ray
energy fluence $F_{\rm tot}$ as
\beq
\label{eiso}
\eiso=4\pi d_L^2F_{\rm tot}/Z,\,\,\,\,\, {\rm where}\,\,\,\,\, d_L=
(c/H_0)Z\int^z_0 dz'/\sqrt{\Omega (z')}
\eeq
is the luminosity distance, and 
$\Omega (z) = \epsilon(z)/\epsilon_0$ describes the evolution of the
total energy density $\epsilon(z)$, of all substance.%
\footnote{Here  $Z\equiv 1+z$, $\epsilon_0=3H_0^2c^2/(8\pi G)$ and 
$\Omega(z)=\Omega_mZ^3+ \Omega_\Lambda$. In what follows we use a
Hubble constant $H_0=70$ km/(s Mpc), and assume a flat universe with
$\Omega_m=0.3$ for matter and $\Omega_\Lambda=0.7$ for the cosmological
constant.}
In this case the cosmological parameters are related to the observations as
\beq
\label{omega}
\left(\int^z_0 {dz'\over \sqrt {\Omega (z')}}\right)^2 = 
\left(H_0\over c\right)^2\left({1\over 4\pi Z}\right)\left({{\cal
E}_0{\cal C}(E_p^{\rm obs}Z/E_0)\over F_{\rm tot}}\right),
\eeq
which involves at least 2 unknown functions $\Omega(z)$ and ${\cal C}(\ep)$, and
3 more  if
the correlation  function ${\cal C}$, and the scales ${\cal E}_0$ and $E_0$
evolve in
time. Without a knowledge of the forms of the evolution functions ${\cal
C}(\ep, z)$, ${\cal
E}_0(z)$ and $E_0(z)$ precise determination of the $\Omega(z)$ is not
possible. 
Most attempts to this end have assumed  not only that the correlation
function has small dispersion (e.g. ${\cal C}(\ep)\propto \delta(\ep-E_0)$)
but also that it remains narrow at all redshifts (i.e. $E_0$ does not evolve),
and that there is no luminosity or $\eiso$ evolution (i.e. ${\cal E}_0$ is a
constant). 

There has been some attempts to determine some aspects of the evolution  (see
e.g Li 2007), but  most of these evolutionary trends have not been addressed.
These
evolutions, in principle, can be determined for an assumed cosmological model
given a large enough sample. However, using such results to determine the
cosmological model
will be a circular and meaningless exercise. Thus, at
this state of our knowledge,
the use of the current  GRB data for determining the global
cosmological parameters seems
premature. We need to  learn more about the nature of the
GRBs and the cosmological evolution of their characteristics before they can be
used for this task. The immediate situation may be more akin
to star forming galaxies and active galactic nuclei (AGNs), whose
characteristics have also broad
distributions and may evolve in time. This fact  has shifted the
focus of activity  (both in galaxies, AGNs and GRBs) to the
investigation of structure formation, the building
process of the black holes,  and  to SFR and CME. 

In recent years there has been increased activity in trying to use
the existing GRB data to determine the shape and evolution of the luminosity
function (LF), $\Psi(L,z)$, the GRB  formation rate density (co-moving)  
${\dot \rho}(z)$ (GRBFR for short), and
its relation to
the SFR or CME (see e.g.  Porciani \& Madau 2001; Natarajan et al. 2005; Daigne
\ea
2006; Jakobson et al. 2006; Le \& Dermer 2007;
Salvaterra et al. 2009; Butler et al. 2010; Wanderman \& Piran
2010; Howell et al. 2014). All these works use the forward fitting (FF)
method, whereby the observed data, such as flux, fluence and redshift
distributions are fit to prediction of models with some assumed
{\it parametric forms} for the numerous  functions [LF, ${\dot \rho}(z)$,
spectrum] and their evolutions.
For example, a
common
practice is to ignore luminosity evolution (see, however, Salvaterra et al.
2012) and assume a GRBFR similar to the
SFR, a power-law LF
with breaks, Band spectrum with unique values
of the high and low energy indexes, etc. A more
objective procedure
would be to determine these characteristics from the data directly and, as much
as possible, {\it
non-parametrically}.  The statistical methods
developed by Efron and Petrosian are designed exactly for this kind of analysis
and have been
used for
analysis of cosmological evolutions of quasars (Maloney \& Petrosian,
1999; Singal et al. 2011; Singal et al. 2013), GRBs (Lloyd et al.
1999); Kocevski \& Liang 2006; Dainotti et al. 2013)
and  blazars (Singal et al. 2012; Singal et al.
2014).%
\footnote{Codes for application of these
methods can be found at \url{http://www.inside-r.org/node/99623} and 
\url{cran.r-project.org/web/packages/DTDA/DTDA.pdf}.}

In this paper we apply these method to data from {\it Swift} on long GRB with
known redshifts with the
aim
of determining the cosmological evolution of the general LF which means the
luminosity and formation rate variation with redshift. This parallels very
closely with an earlier work using pre and early {\it Swift} data which formed
a chapter of Aurelien Bouvier's PhD thesis at Stanford (Bouvier 2010). The
results of this work can also be found in Petrosian et al. (2013). In the next
section we
describe the data we use and in \S 3 we present a brief description of the
method as applied to these data. The results are presented in \S 4 and a brief
summary and discussion is given in \S 5.

\section{Swift Data and Selection Effects}

\begin{figure}[!h]
\begin{center}
\includegraphics[width=15cm]{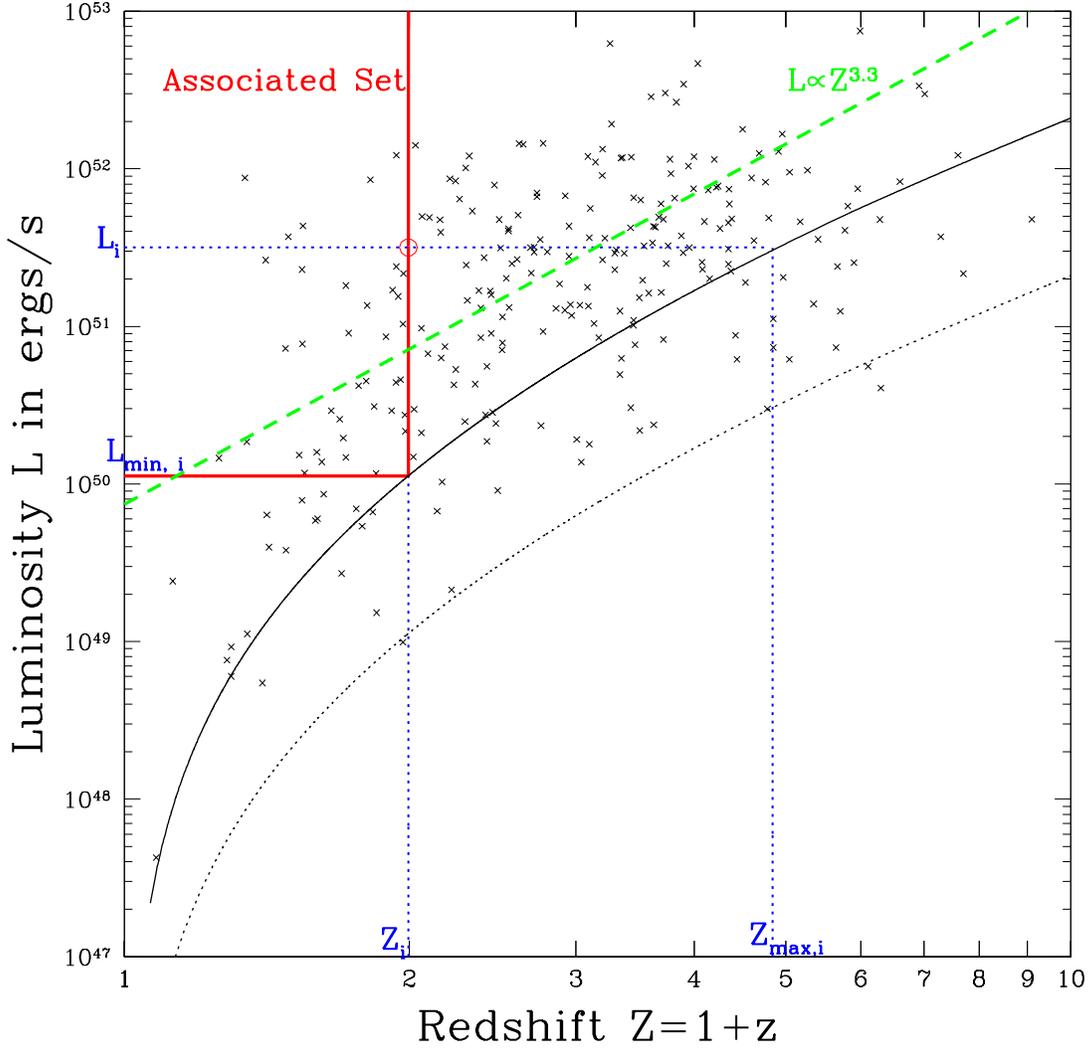}
\caption{K-corrected Luminosity vs. redshift. The solid and dotted (black)
curve shows
the truncation due to a flux limits of $f_{\rm lim}=2\times 10^{-8}$ and
$2\times 10^{-9}$ erg/(s cm$^2$), respectively. In our analysis we use the
larger and more conservative limit.  The dashed (green) line shows the best fit
luminosity
evolution to the
raw
data (data points above the solid curve. Most of this correlation  is due to the
truncation.
The solid (red) vertical and horizontal lines define the boundaries of the
associated
set (in this case $M_i$) of the source marked by the red circle. The 
dotted (blue)
lines and letters show the luminosity and redshift of the source and  the
maximum redshift and the minimum luminosity,
defined in Eq.~(\ref{limits}), that this source can have and still be in the
sample.}
\label{LvsZ}
\end{center}
\end{figure}

Over the last decade,  the Burst Alert Telescope (BAT) on board the {\it Swift}
satellite has
detected more than 800 long GRBs with peak fluxes in the 15-150 $\unit{keV}$
energy band and above the threshold flux $f_{\rm p, lim}\sim 2\times 10^{-8}$
erg/(s cm$^2$). Approximately,  90\% by the {\it X-ray Telescope} (XRT)
and about one third 
by both XRT and the {\it Ultraviolet/Optical Telescope} (UVOT). Redshifts
obtained
for over 250
{\it Swift} GRBs  from these instruments and the follow up observations they
enable on larger,
ground-based telescopes. Fig.~\ref{LvsZ} shows scatter diagram of luminosities
and redshifts for 253 long GRBs, taken
from Nysewander et al. (2007), Butler et al. (2009), and  NASA's online burst
archive, which collates data from Evans et al. (2009) and the Gamma-ray
Coordination Network circulars. 


From the redshifts and the observed peak gamma-ray energy flux $f_p$, integrated
over the observed \swiftA-band (15--150 keV), we
calculate the peak luminosity for the assumed cosmological model 
and K-corrected  for the same rest frame energy band as%
\footnote{Note that this definition of the K-correction follows the original
definition that can be found in e.g. Peebles (1993). Sometimes the inverse of
this is use relating a fixed rest frame energy band to a variable observed
band, e.g Bloom et al. (2001) in which case $K(Z)=\frac
{\int_{15\unit{keV}}^{150\unit{keV}}E{f(E)}\,\mathrm{d}E}
{\int_{15\unit{keV}/Z}^{150\unit{keV}/Z}E{f(E)}\,\mathrm{d}E}.$ For a power law
spectrum these give identical results but could be different when there are
significant spectral deviation from a simple power law.}
\beq\label{lum}
L=4\pi d_L^2(z,{\Omega})f/K\,\,\,\ {\rm with}\,\,\,\,K(Z)=\frac
{\int_{15Z\unit{keV}}^{150Z\unit{keV}}E{f(E)}\,\mathrm{d}E}
{\int_{15\unit{keV}}^{150\unit{keV}}E{f(E)}\,\mathrm{d}E}.
\eeq
The spectra  $f(E)$ are fitted to
either a power law $f(E)=A(\frac{E}{100})^{\alpha}$, a power law with an
exponential cutoff
$f(E)=A(\frac{E}{100})^{\alpha}{e}^{-\frac{(2+\alpha)E}{E_{p}}}$, or the Band
model
\bea\label{band}
[f(E)= \begin{cases} 
      A(\frac{E}{100})^{\alpha}{e}^{-\frac{(2+\alpha)E}{E_{p}}} & E<E_{br}\equiv \frac{(\alpha-\beta)E_{p}}{2+\alpha} \\
      A(\frac{E}{100})^{\beta}{e}^{\beta-\alpha}(\frac{E_{br}}{100})^{\alpha-\beta} & E \ge E_{br} 
   \end{cases} 
\eea

In Figure~\ref{LvsZ} the dotted and solid (black) curves show 
the truncation $L>L_{\rm min}(z)=4\pi d_L^2(z,{\Omega})f_{\rm
min}/{\bar K}$ due to two {\it fiducial  peak flux thresholds} $f_{\rm lim}$ 
and
for ${\bar K}$ correction calculated for an average  spectrum. To insure a
higher completion level in what follows we use he larger limiting flux of
$f_{\rm lim}=2\times 10^{-8}$ erg/(s cm$^2$) that includes 207 sources. The
smoothed  redshift
distribution of this subsample is shown by the dotted (red) curve in the left
panel of
Fig.~\ref{sigma}.  The
truncation induces a strong luminosity-redshift correlation, $L\propto
Z^{3.3}$, shown by the dashed (green) line, which complicates the
determination
of the true
intrinsic correlation, namely the  luminosity evolution. 

However, in addition to the bias involved in the detection of the prompt
emission, the measurement of the redshift can introduce further truncations.
X-ray and optical afterglows are vital for timely and accurate burst
localization and redshift measurement. It is difficult to quantify the optical
selection criterion. But  as discussed in \S 4.3, we can include in our
analysis the effects of X-ray selections. Since the optical and X-ray
afterglow fluxes show relatively strong correlation, inclusion of X-ray
selection effects may to some degree alleviate the problem of the optical
selection effects.

\section{Methods and Approach}\label{method}

The problem under consideration here requires determination of multi variate
distribution from data truncated by observational selection effects. We will
demonstrate our approach for
determination of single LF and its evolution;  $\Psi(L,z)$ from a flux limited
sample such as that shown in Fig.~\ref{LvsZ}. Without loss of generality, we
can write the LF as 
\begin{equation}
\label{lf}
\Psi(L,z) = {\dot \rho} (z) \psi[L/g(z),\alpha_i] / g(z),
\end{equation}
where, in addition to the GRBFR ${\dot \rho}(z)$, we have introduced $g(z)$ and 
$\alpha_i$ to describe  luminosity and shape evolutions,
respectively. Given sufficiently  large sample one can determine all three
evolutions. Our experience with evolution of the LF in other extragalactic
sources such as quasars and blazars indicates that the least variable among
these
is the shape parameter(s). Thus, because of the
limitations of the current GRB data we will focus here only on the GRBFR
and
luminosity evolutions and assume  constant shape parameters. The truncation
or bias introduced by the flux limit [$f>f_{\rm lim}$ or $L>L_{\rm  min}(z)=4\pi
d_L^2(z)f_{\rm lim}/K(z)$] is known as the {\bf
Malmquist Bias}. Many papers have been written to correct for this bias in
astronomical literature (Malmquist 1925); Eddington 1940; Trumpler \&
Weaver 1953; Neymann \& Scott 1959).  Most of these
early procedures assumed simple parametric forms (e.g Gaussian) distributions. 
However, since the discovery of quasars  the tendency has been to use
non-parametric methods: e.g $\langle V/V_{\rm max}\rangle$
(Schmidt 1972; Petrosian 1973) or the $C^{-}$
(Lynden-Bell 1971). For a more detailed description see review
by Petrosian (1992). All these methods however, suffer from a
major shortcoming because they assume that the variables
are \underline{independent or uncorrelated},  or  that the LF
is separable; $\Psi(L,z)=\psi(L){\dot \rho}(z)$. This ignores luminosity
evolution and is an assumption that is made commonly  for GRBs which  turns
out to be  incorrect (see below). 

Thus, the first task should be the determination of the correlation between the
variables. Unfortunately, most past works dealing with determination
of the GRB LF or GRBFR (those mentioned above and others; e.g. Kistler et al.
2007; Li 2008) omit this crucial step, which, as described below can lead to a
incorrect GRBFR.
A commonly used non-parametric method for determining correlations  is  the
Spearman Rank
test. This method,
however, fails for  truncated data. Efron \& Petrosian (1992)  developed a novel
method to account for this truncation. The gist of the method is to
determine the rank  $R_i$ of each data point $L_i$ (or $z_i$) among the $N_i$
(or $M_i$)  members of its {\it associated set}, or its largest un-truncated
subset. The region bounded by the red
lines in Fig.~\ref{LvsZ} depicts the boundary of this set for the data set
$L_i,z_i$ identified by the red circle. This is for ranking in redshifts (from
low to high values) and the set includes $M_i$ sources with $z_j<z_i$ and
$z_{{\rm max},j}>z_i$ (or $L_j>L{{\rm min},i})$. In an analogous manner one can
define the set of $N_i$ sources with 
$L_j>L_i$ and $L_{{\rm min},j}(z)<L_i$ (or $z_j<z_{{\rm max},i}$ for ranking the
luminosities from above. These limiting values, also shown in Fig.~\ref{LvsZ},
are given as 
\beq
\label{limits}
L_{{\rm min},j}=4\pi d_L^2(z_j)f_{\rm lim}/{\bar K}(z_j)\,\,\,\,{\rm and}
\,\,\,\,L_j(z)=4\pi d_L^2(z_{{\rm max},j})f_{\rm lim}/{\bar K}(z_{{\rm max},j}).
\eeq
Then using a test statistic, e.g. Kendell's
tau defined as $\tau=\Sigma_i(R_j-E_j)/\sqrt{\Sigma_i V_j}$, one can determine
the degree of correlation. Here $E_j$ and $V_j$ are the expectation and variance
of the ranks.  A small value ($\tau\ll 1$) would imply independence, and
$\tau>1$ would imply significant correlation. In the latter case one can
redefine the variables, e.g. define a ``local" luminosity $L'=L/g(z)$ [with
$g(0)=1$] 
using a parametric form, e.g.  $g(z)=(1+z)^k$, and calculate $\tau$
as a function of the parameter $k$.%
\footnote{Note that for each $k$ the limiting values $L_{{\rm min},j} (z_{{\rm
max},j})$ and associated sets $N_j (M_j)$ are calculated anew.}
 The values of $k$ for which
$\tau=0$ and  $\tau=\pm 1$ give the best value and one sigma range for
independence.  

Once independence is established then one can use the above mentioned 
non-parametric
methods ($\langle V/V_{\rm max}\rangle$  or the
$C^{-}$) to
determine the mono-variate distributions; local LF $\psi(L')$ and the formation
rate evolution ${\dot \rho}(z)$. We will use the $C^-$ method which  builds  up
the cumulative 
distributions of both variables $L$ and $z$ defined as 
\beq\label{cums}
{{\dot \sigma}}(z)=\int_0^z{{\dot \rho}(z')\over
Z'}{dV(z')\over dz'}dz'\,\,\,\,{\rm 
and}\,\,\,\,
\phi(L')=\int_{L'}^\infty \psi(x)dx
\eeq
point by point non-parametrically, again
using the
concept of the {\it associated set}. For example
$\phi(L_i)=\prod_{j=2}^i(1+1/N_j)$. The cumulative
functions can then be differentiated to get $\psi$ and ${\dot \rho}$ using the
derivative of the  $V(z)$, the co-moving volume up to $z$. 

The observed redshift
distribution is
related to the cumulative functions as 
\beq\label{dNdz}
{dN\over dz}={d{\dot \sigma}(z)\over dz}\phi[L'_{\rm min}(z)]\,\,\,\,{\rm
with}\,\,\,\, L'_{\rm min}(z)={4\pi d_L^2(z)f_{\rm min}\over {\bar K(z)}g(z)},
\eeq
where $\phi[L'_{\rm min}(z)]$ represents the effects of the truncation
and $d{\dot \sigma}(z)/dz$ gives the true redshift distribution of the
parent population; i.e. what
one would observe in the absence of truncation (i.e. when flux thresholds
$f_{\rm lim} \rightarrow 0$).

It should also be noted that, this method is
not limited
to a simple  flux limited data but, as shown in Efron \& Petrosian
(1999), it can deal with truncations from both below and above, or  with
the most general truncation situation
where each data point, say [$L_i, z_i$], has its individual upper and lower
limits, [$L_{i,{\rm min}}<L_i<L_{i,{\rm max}}$ and $z_{i,{\rm
min}}<z_i<z_{i,{\rm max}}$]. We will not be dealing with such complications
here, since our data is truncated from below only.%
\footnote{ We note, however, that  this
is an enormous advantage because it allows one to combine data
from many different regions of sky with different backgrounds and even from
different instruments obtained with different selection criteria.}

In appendix A we apply this method to a  flux limited sample selected
from a parent simulated sample with known charcteristics of the LF and 
evolution, and demonstrate that we recover the input characteristics 
accurately.

\begin{figure}[!h]
\begin{center}
\includegraphics[width=10cm]{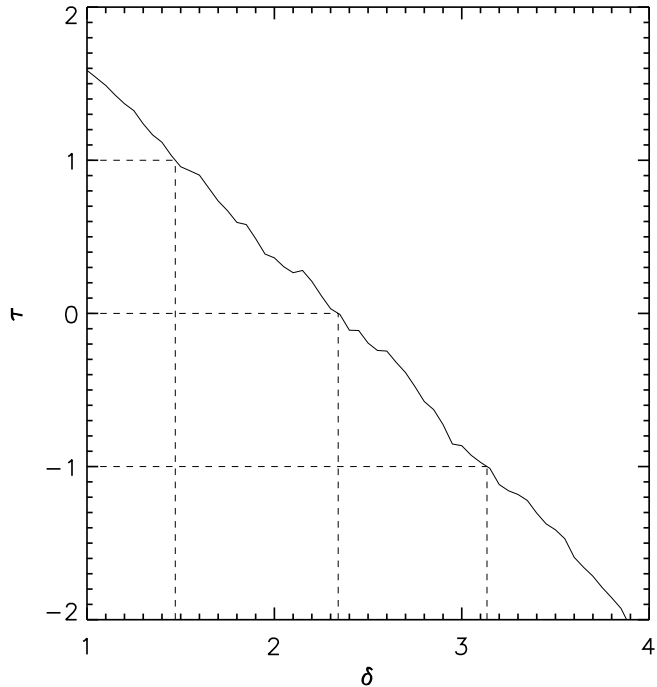}
\caption{Variation of Kendell's $\tau$ statistics with the evolution index
$\delta$ showing independence for $\delta=2.33$ with one sigma range 1.48 to
3.12. We set $\delta'=\delta$ and $Z_{\rm cr}=3.5$ in Eq.~(\ref{lumevol}).}
\label{tau}
\end{center}
\end{figure}

\section{Results}\label{sec:results}

\subsection{Luminosity Evolution}\label{sec:Levol}

As stressed above, the first task
is to carry out test of the independence of $L$ and $z$ and determine the
luminosity evolution. As shown above the raw data shows strong correlation
$L\propto Z^{3.9}$. 
Using the associated sets to account for the truncations we find a Kendell's
$\tau\sim 3.5$ indicating the presence of a strong intrinsic correlation (i.e.
luminosity evolution). The form of the evolution can be determined
parametrically. The form often used is $g(z)=Z^\delta$ which is appropriate for
low and intermediate redshifts. But at higher redshifts this becomes excessively
large considering the fact that the time intervals
$dt=dZ/(Z\sqrt{\Omega_MZ^3 + 1 - \Omega_M})$ decreases rapidly with increasing
redshifts for $Z > 3$. Based on our experience in treating the evolution of
the quasars we use the following, that has a slower luminosity evolution for
$Z>Z_{\rm cr}$ 
\beq\label{lumevol}
g(z)={Z^{\delta}\times Z_{\rm cr}^{\delta'} \over Z^{\delta'} + Z_{\rm
cr}^{\delta'}}.
\eeq
After some trial and error we settled on $Z_{\rm cr}=3.5$ and
$\delta'=\delta$, leaving us with  one free parameter. We then define the
``local" luminosity $L'=L/g(z)$ (and the truncation curve $L'_{\rm
min}(z)=L_{\rm min}(z)/g(z)$) and calculate the statistic $\tau$ as  a function
of $\delta$. Fig.\,\ref{tau} shows the results indicating independence of $L'$
and $z$ for $\delta=2.3$ with the one sigma range $[1.5 - 3.1]$. This is
significantly less than the raw index $\sim 4$, but
still considerable evolution; factors of $4.0, 7.8$ and $\sim 11$ at
redshifts 1, 2  and 3, respectively. This is in agreement with recent
results using FF methods (e.g. Salvaterra et al. 2012) and 
earlier results using the Efron-Petrosian method on pseudo-redshift samples
(Lloyd-Ronning et al. 2002; Yonetoku et al. 2004;  Kocevski \& Liang 2006) and
samples with redshift (Petrosian et al. 2013). 

\subsection{The Luminosity Function}\label{sec:LF}

Transferring the observed $L-z$ scatter diagram to the local
luminosity $L'-z$ diagram 
we can
now determine the local LF $\psi(L')$ using the $C^-$ method.
Since we have assumed that the shape of the LF is invariant this function 
when shifted in luminosity by $g(z)$ describes the LF at all redshifts. The
left panel of Fig.~\ref{cums}
shows the 
cumulative local LF $\phi(L')$ build up point by point starting with the
highest
observed luminosity (filled points). For comparison we also show the raw
cumulative distribution (open circles)  showing the  correction  due  to
truncation
obtained by our method. As
shown in Appendix B one can also obtain a histogram of the differential LF
\beq
\label{diffLF}
\psi(L')=-{d\phi(L')\over dL'}=-{\phi(L')\over L'}{d\log \phi(L')\over d\log L'}
\eeq
directly from the data. However,
because of the paucity of the data and the
random nature of the luminosities, their separation
(e.g $\delta \log L_i=\log (L_{i-1}/L_i)$)
has a large dispersion yielding a noisy differential LF. Instead we obtain the
differential LF by taking the 
derivative of appropriately smoothed curve fitted to $\phi(L')$. As shown by
the dashed (red) and dotted (green) curves in Fig.~\ref{cums} (left) a smooth
broken power law or a power law with exponential cutoff provide adequate fit to
the cumulative LF. Derivatives of this give a similar description for the
differential LF. For example the  indicies -0.5 and -2.2  of $\phi(L')$ means
that the differential luminosity function can also be
fitted to a broken law with indicies  $d\log \psi/d\log L'\sim -1.5$ and $-3.2$.
This is  a steeper LF, especially at the high end,
than some recent results based on FF method; e.g. Howell et al. (2014)   obtain
indexes $[-0.95, -2.59]$, and Salvattera
et al. (2012) obtain indexes $[-1.5, -2.3]$.

\begin{figure}[!h]
\begin{center}
\includegraphics[width=8.0cm]{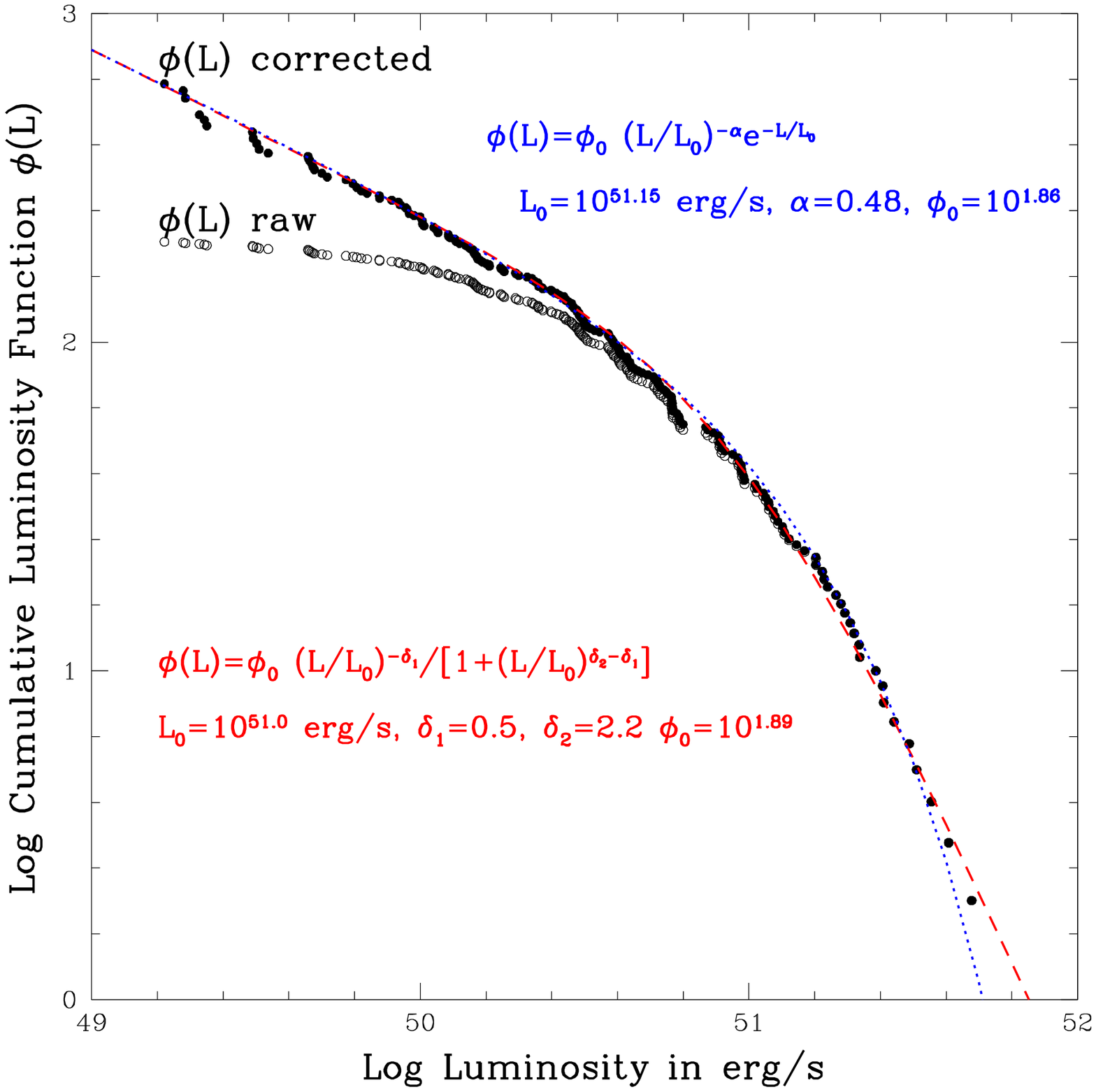}
\includegraphics[width=8.0cm]{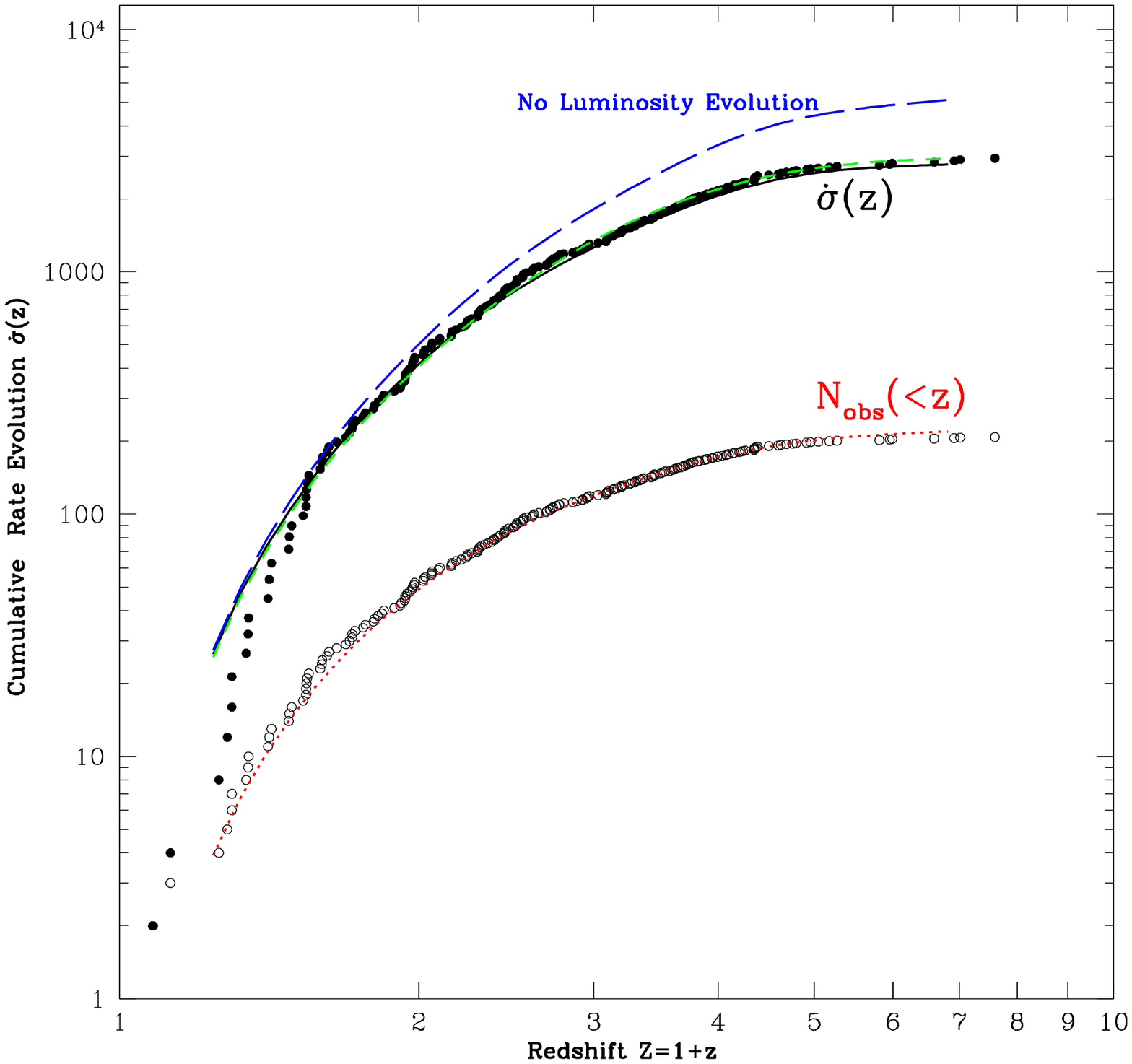}
\caption{{\bf Left:} The cumulative 
LF. The filled
points shows the cumulative local LF $\phi(L')$ while the open
points shows the cumulative distribution obtained ignoring the
effects of the truncation. We have omitted 5 GRBs with $L<10^{49}$ erg/s.
The  dashed (red) curve shows a broken power law fit with indicies -0.5
and -2.2 above and below $L_0=10^{51}$ erg/s and the dotted (blue) curve shows
a fit to Schechter function; a power-law with exponential cutoff.
{\bf Right:} The cumulative rate evolution ${\dot \sigma}(z)$ versus
redshifts 
corrected for selection effects due to the $\gamma$-ray flux limit: 
filled points with solid (black) line a smoothed fit. The very similar dashed
(green) curve is obtained using both
$gamma$ and X-ray threshold effects. The long-dashed (blue) is obtained ignoring
luminosity evolution.
The open points, and fitted dotted (red) curve, shows the observed cumulative
distribution $N(<z)$,
which of course does not include
effects of the truncation. }
\label{cums}
\end{center}
\end{figure}

\subsection{Formation Rate Evolution}\label{sec:rho}

\begin{figure}[!h]
\begin{center}
\includegraphics[width=8.0cm]{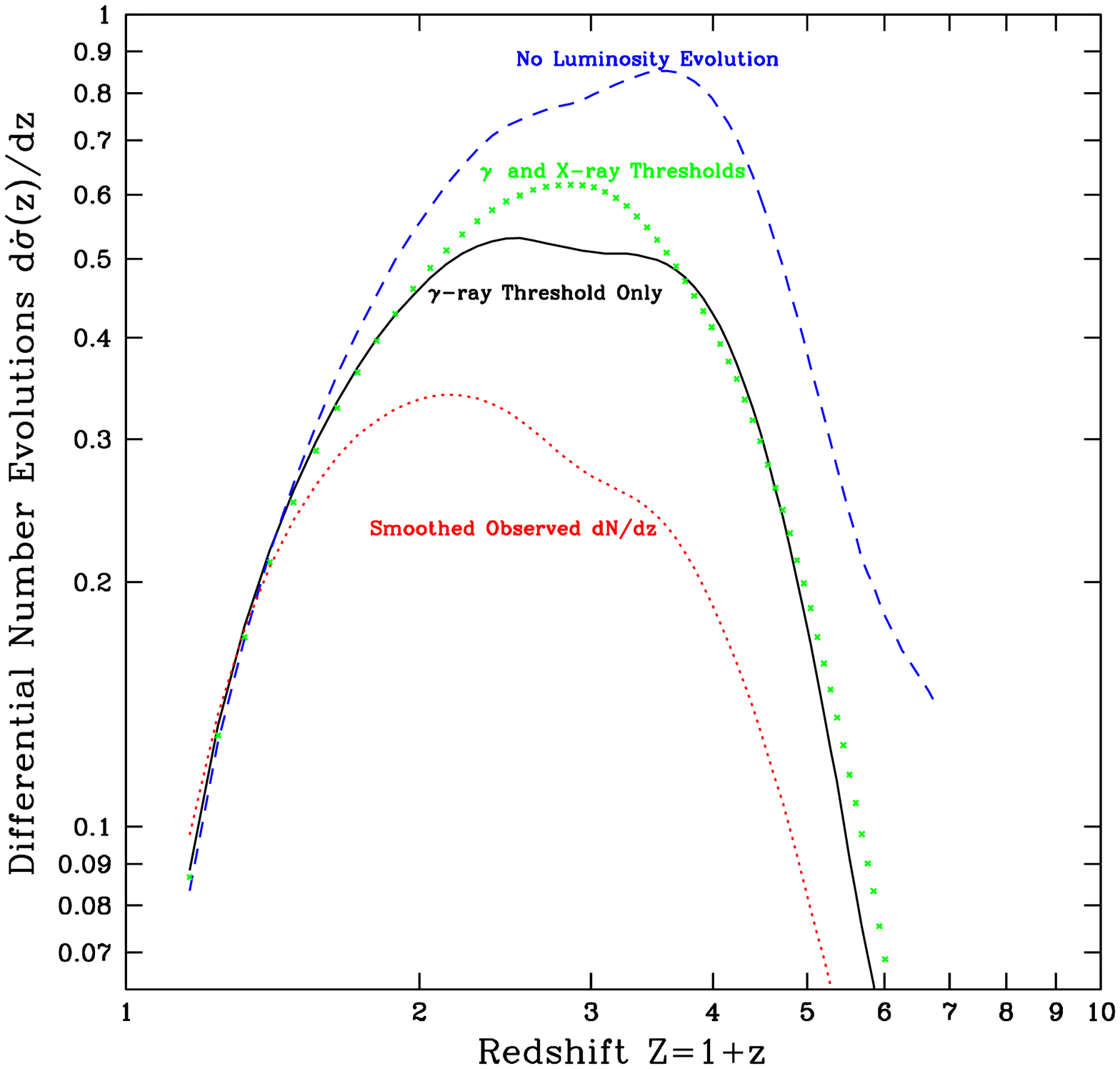}
\includegraphics[width=8.0cm]{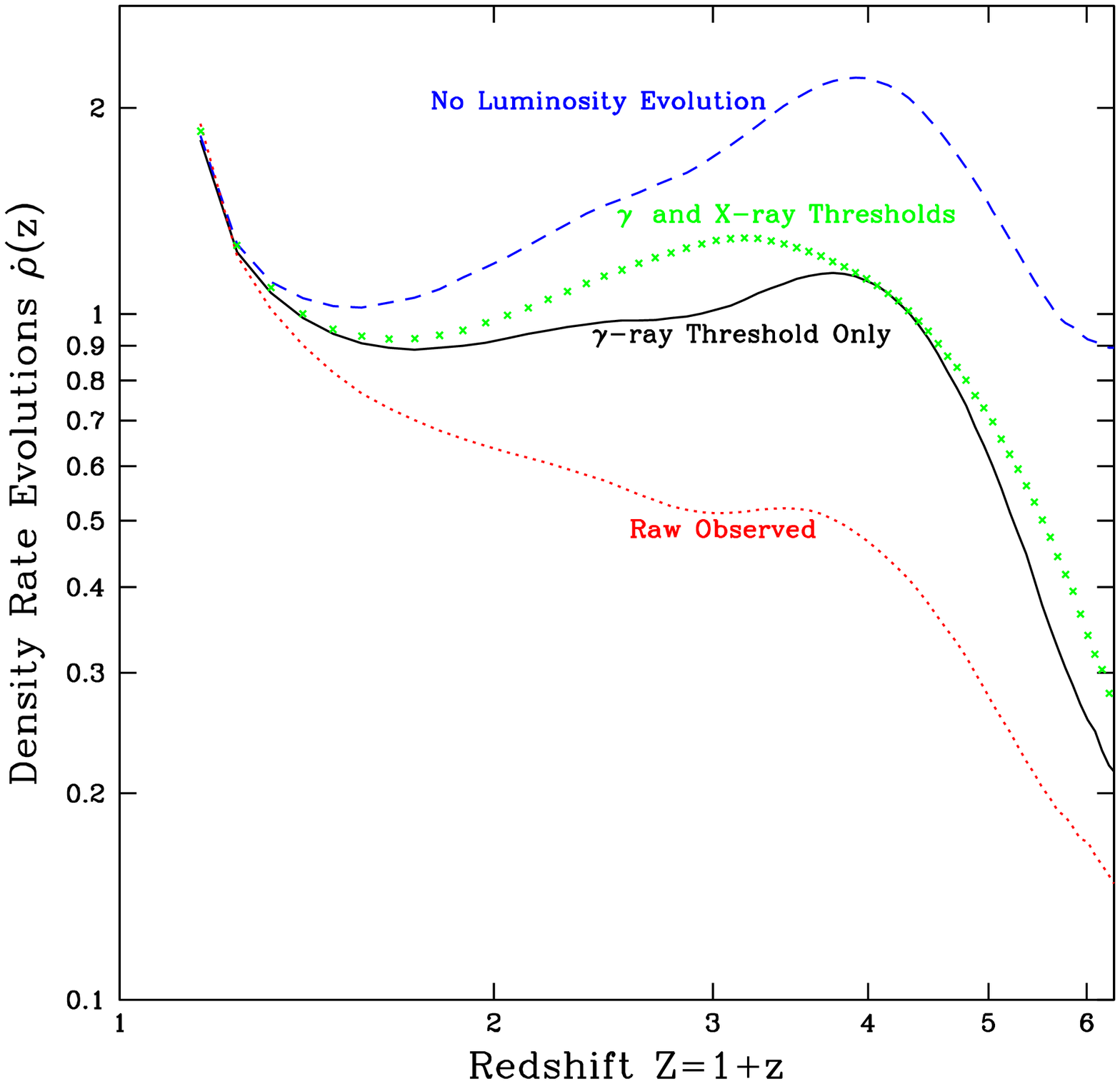}
\caption{{\bf Left:} Comparison of the observed redshift 
distribution $dN/dz$ (dotted, red) with the intrinsic redshift distributions
$d{\dot \sigma}/dz$ (solid, black) obtained from the
differentiation of curves fitted to the cumulative distribution shown on the
right panel of Fig.~\ref{cums}. The dashed (blue) curve shows the derived
redshift distribution ignoring the luminosity evolution demonstrating the
importance of inclusion of the luminosity evolution. The x points (green)
are obtained including the
effects
of both gamma-ray and X-ray selection biases, as well of the luminosity
evolution, as described in the text. This is
very close to the black curve indicating that most of the selection bias
is already accounted for by the correction for the gamma-ray threshold.
These  curves are normalized to redshifts $z<1$ where the observational
selection bias is expected to be small. 
{\bf Right:} The density rate evolution obtained from the
distributions on the left and Eq.~(\ref{rho}) using the same color codes.
The dashed (red) curve is obtained using the smoothed
observed $dN/dz$ which shows a much more rapid decline than the true rate 
obtained for $\gamma$-ray threshold only (solid black),  both 
$\gamma$ and X-ray thresholds (long-dash green). The dashed (blue)
curve is obtained ignoring the luminosity evolution. 
All rates are normalized at $z<1$.
}
\label{sigma}
\end{center}
\end{figure}

From the $L'-z$ scatter diagram we can also obtain the GRBFR
evolution.
The   cumulative rate evolution ${\dot \sigma}(z)$ is shown by the filled points
on the
right panel of  Fig.~\ref{cums}. This is obtained point by point starting
with the lowest redshift. Again, for comparison  we also show (open points) the
raw observed 
cumulative distribution $N(<z)=\int_0^z(dN/dz)dz$. On the left panel of
Fig.~\ref{sigma} we compare the smoothed differential observed redshift
distribution
$dN/dz$ (dotted red) with the true  distribution $d{\dot
\sigma}(z)/dz$
(solid, black), obtained from the derivative
of a smooth solid (black) curve  fitted to ${\dot \sigma}(z)$ on Fig.~\ref{cums}
(right).  Both of the above 
comparisons show that observations miss many intermediate and high
redshift sources due
to truncations
affected by the gamma-ray threshold.  These curves are normalized at low
redshifts where the effect of truncation is expected to be minimal. A 
proper normalization  can be carried out using
total population counts like the so-called log$N$-log$S$ diagram, which is
beyond the scope of his
paper.

The co-moving GRBFR evolution is obtained  as 
\beq\label{rho}
{\dot \rho}(z)=Z{d{\dot \sigma}(z)/dz\over dV(z)/dz}.
\eeq
This is shown by the solid (black) curve on the right panel of Fig. \ref{sigma}.
In this figure we
also show the raw GRBFR   that one would obtain ignoring
the
truncation (dotted, red), i.e. using the observed $dN/dz$ instead of
$d{\dot \sigma}(z)/dz$
in Eq.~(\ref{rho}). As expected this gives much lower rate at higher
redshifts. We have also calculated GRBFR evolution 
ignoring the  luminosity evolution, i.e.
determining the distributions from the $L-z$ data shown in Fig.~\ref{LvsZ}
instead of $L'-z$ scatter diagram.
This is shown
by the dashed (blue) curves in Figs.~\ref{cums} and \ref{sigma}. As evident
ignoring
the luminosity evolution overestimates the required GRBFR. This is the reason
that some of earlier works (cited above) that ignored luminosity evolution
obtained a high GRBFR at high redshifts, higher than the observed SFR. Again we
have normalized these  rates at low redshifts ($z<1$) where the observational
selection effects are smallest.   

All of the above results are obtained including only the effects
truncation due to prompt $\gamma$-ray detection threshold. As mentioned
in \S 2,
other observational selection effects, arising in the process of localization
and securing redshifts, add more truncations. In particular the
threshold of X-ray detection, the first crucial step for determination of the
redshift, is an important factor. Based on the X-ray flux data from
Racusin  et al. (2011; private communications), we  use a conservative limit of
$\sim 2\times 10^{-13}$ erg/(s cm$^2$) (in the 0.2-10
keV range and at 11 hrs after the trigger) 
to determine how
this additional selection bias affects our result. To this end we have
repeated the above
calculations using
the effects of both gamma-ray and X-ray thresholds. This extra truncation
redefines 
the {\it associated sets} which, as described above, for a single threshold
consists of the set with
$z_j<z_i$ and $z_{{\rm max},j}<z_i$, where $z_{{\rm max},j}$, defined in
Eq.~(\ref{limits}), is the maximum
redshift the source with $L_j, z_j$ can be moved to and  still be in the sample.
With two thresholds we now have two limiting redshifts. We use the minimum of
the two to define the associated sets.   The x points (green)  on both
panels of
Figs.~\ref{sigma}  show this results. These are very similar to the black
curves obtained with the $\gamma$-ray threshold effects alone. This is
encouraging because it may imply that the other selection effects in securing a
redshift, which cannot be easily quantified, have also small effects.
Comparing these with the raw (uncorrected)  redshift distributions
(dotted red curve) we conclude that  most of the
correction has already been accounted for by the gamma-ray threshold.
We also note that, there is a strong
correlation between X-ray and
optical fluxes of \swift GRBs which
indicates that the X-ray threshold may be good proxy for the optical
selection effects. 

Finally we note that a
more rigorous
approach to the multivariate nature of this problem would require us to obtain
the evolution of the bivariate LF
$\Psi(L_{\gamma},L_{X},z)$. As demonstrated in Singal et al. (2011) and
(2013), the methods
used in this paper for  a single luminosity can be generalized to higher
dimensions without much
difficulty.  Similarly, if and when  we  understand the optical
selection criterion, we can
include optical luminosities as well in this multivariate distribution.

\begin{figure}[!h]
\begin{center}
\includegraphics[width=14cm]{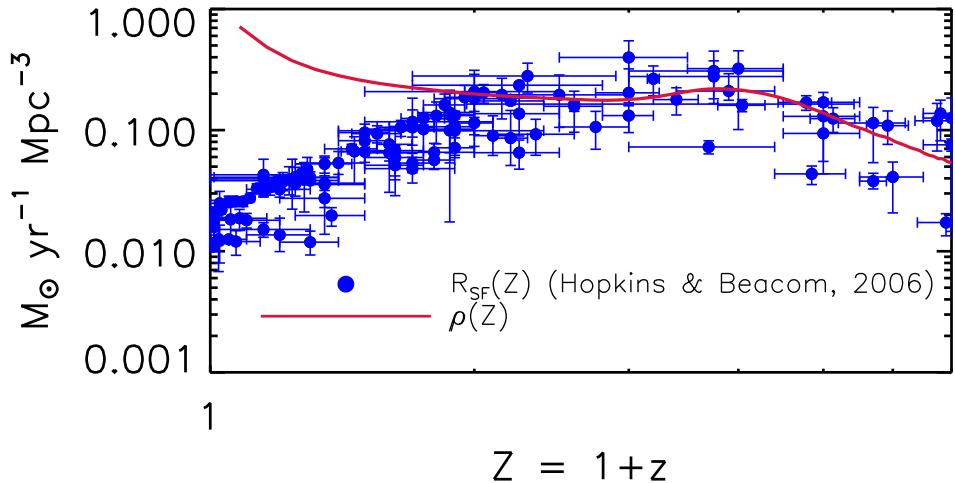}
\caption{Comparison of the the derived GRBFR (dash-dotted, green curve on
Fig.~\ref{sigma}, right) and the SFR from Hopkins \& Beacom (2006) normalized
around  redshift $z\sim 3$. The GRB  rate (red) agrees remarkably well in the
somewhat flat
portion of the SFR rate and decays in a  similar way at high redshifts.
But at low redshifts it has a dramatically different shape and much higher
rate.}
\label{SFR}
\end{center}
\end{figure}

\section{Summary and Discussion}

We have analyzed a sample of about 200 {\it Swift} flux limited long GRBs with
known
redshift using the non-parametric methods of Efron \& Petrosian to obtain the
general evolution of the luminosity function. Our results can be summarized as
follows:

\begin{itemize}

\item

We find that the observed strong  correlation between the luminosity and
redshift is not totally due to  the effects of the flux limit but that there is
significant intrinsic correlation indicating that GRBs have undergone a strong
luminosity evolution at least up to  $z\leq 3$ which can be approximated as
$L\propto (1+z)^{2.3}$.

\item

Correcting for this evolution we then obtain the local LF which can be
represented by a broken power law or a Schechter type function.

\item

We also find the true redshift distribution and the (co-moving) density rate
evolution which starts at a maximum level at lowest redshifts, declines
slowly up to $z\sim 1$ and merges into  a plateau  or an almost  constant rate
between
redshift 1 and 3,  and then shows a relatively
steep decline at higher redshifts. 

\item

Thus, we find that GRBs undergo both luminosity and density evolution; they were
more luminous but less numerous in the past than today. Often the luminosity
evolution is ignored which as shown here gives an incorrect (higher) density
rate
evolution at high redshifts. This is the source of many claims that GRBFR is
higher than SFR at higher redshifts  (see,
e.g. Kistler et al. 2008; Salvaterra et al. 2012; Wei et al. 2014). Robertson
\& Ellis (2012), and Dainotti et al. (2015) using afterglow emission, find a
slower GRBR evolution, but their result also
disagrees with ours because of their neglect of luminosity evolution.

\item

Most analyses of GRB evolution assume a formation rate similar to the SFR
especially at low redshift, but with an increase rate at higher $z$'s. The
additional evolution is
derived by FF to the observed
distribution of redshifts and fluxes.  
With our non-parametric method we obtain directly the GRB formation rate which
is quite different than those derived by FF methods. As shown in the right panel
of Fig.~\ref{SFR}, our derived rate agrees very well with the SFR at redshifts
$z>1$, in particular in the plateau range $1<z<3$, and shows a similar rapid
decline at higher redshifts. But at low redshifts $z<1$ the GRB rate deviates
by more than a decade from the SFR. This is puzzling because,
according to the
prevailing view that GRB formation is favored in low metalicity regions,  one
would expect a lower GRB rate in metal rich  lower redshift galaxies.

\end{itemize}

There are many potential factors that can account for this discrepancy. The
above results are based on the LF of the prompt emission and includes
rigorously only the $\gamma$-ray threshold effects. But the data used involved
also X-ray, UV and optical (space and ground based) measurement for localization
and securing redshifts. We have included the X-ray  detection threshold as best
as can be done and find  a small effect.  The discrepancy at low redshifts
remains large. It appears that the $\gamma$-ray
threshold effects account for most of the data truncations. This is encouraging
because it may indicate that less quantifiable UV and optical selection biases
may have similarly small effects. It should be noted that these added selection
effects are expected to be more important at high redshifts so that they, most
likely, are not the source of the discrepancy at low redshifts. The fact that
all curves in Fig.~\ref{sigma} have same shape at low redshifts supports this
view.

Another possible factor may be the paucity of low redshift and low luminosity
GRBs which dominate the shape of the nearby GRB rate. There have been some
claims that low luminosity GRBs may belong to  a separate class. However, as
can be seen from Fig.~\ref{LvsZ},  eliminating GRBs
with  $L_{\rm min}\leq 10^{49}$ erg/s from the list will affect only the result
at $z<0.4$ leaving a large discrepancy for $0.4<z<1$. This discrepancy,
therefore, may have an important
implication on the formation of long GRBs.

Acknowledgements: This work is partially supported by {\it Swift} guest
investigator grants (NASA  NNX12AE74G) and is based on Ellie  Kitanidis's
senior honor
thesis at Stanford
University (see \url{http://purl.stanford.edu/xp981bq5003}).

\section*{Appendix A: Testing the Methods}

\begin{figure}[!h]
\begin{center}
\includegraphics[width=7.5cm]{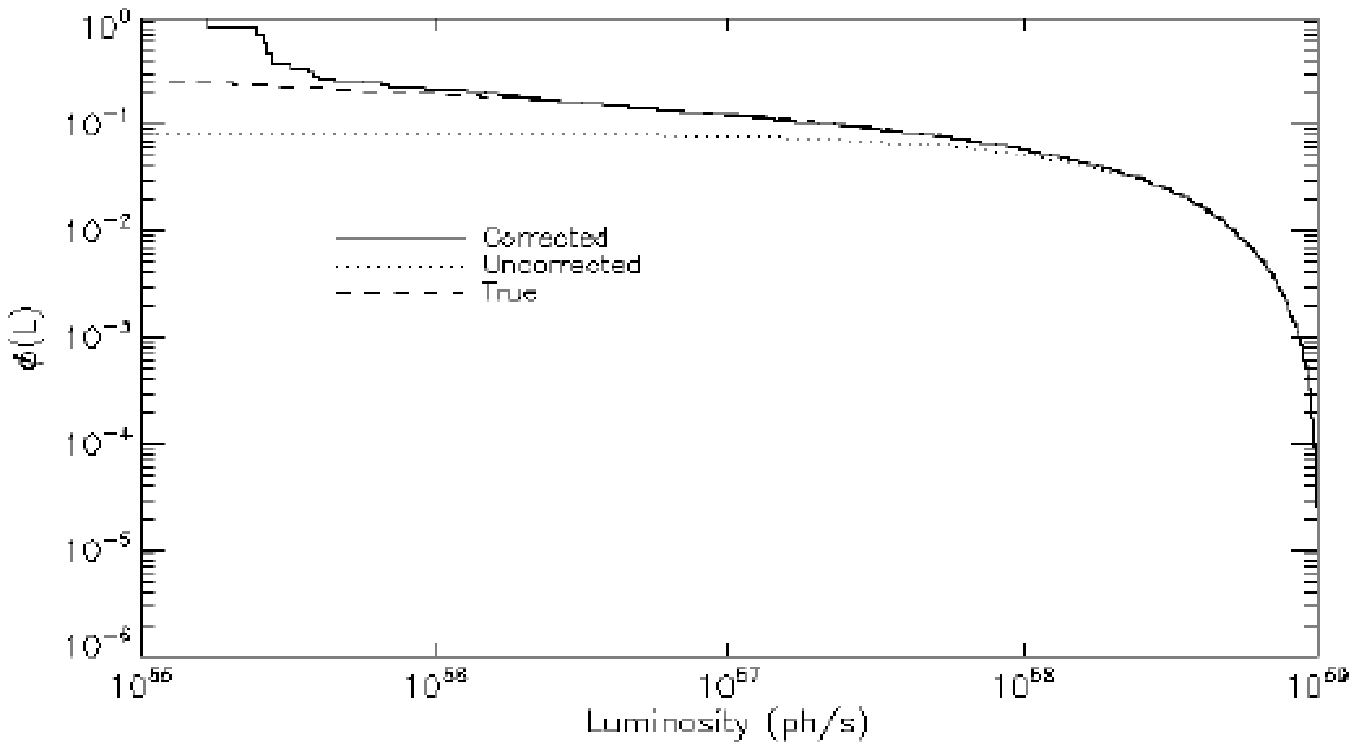}
\includegraphics[width=7.5cm]{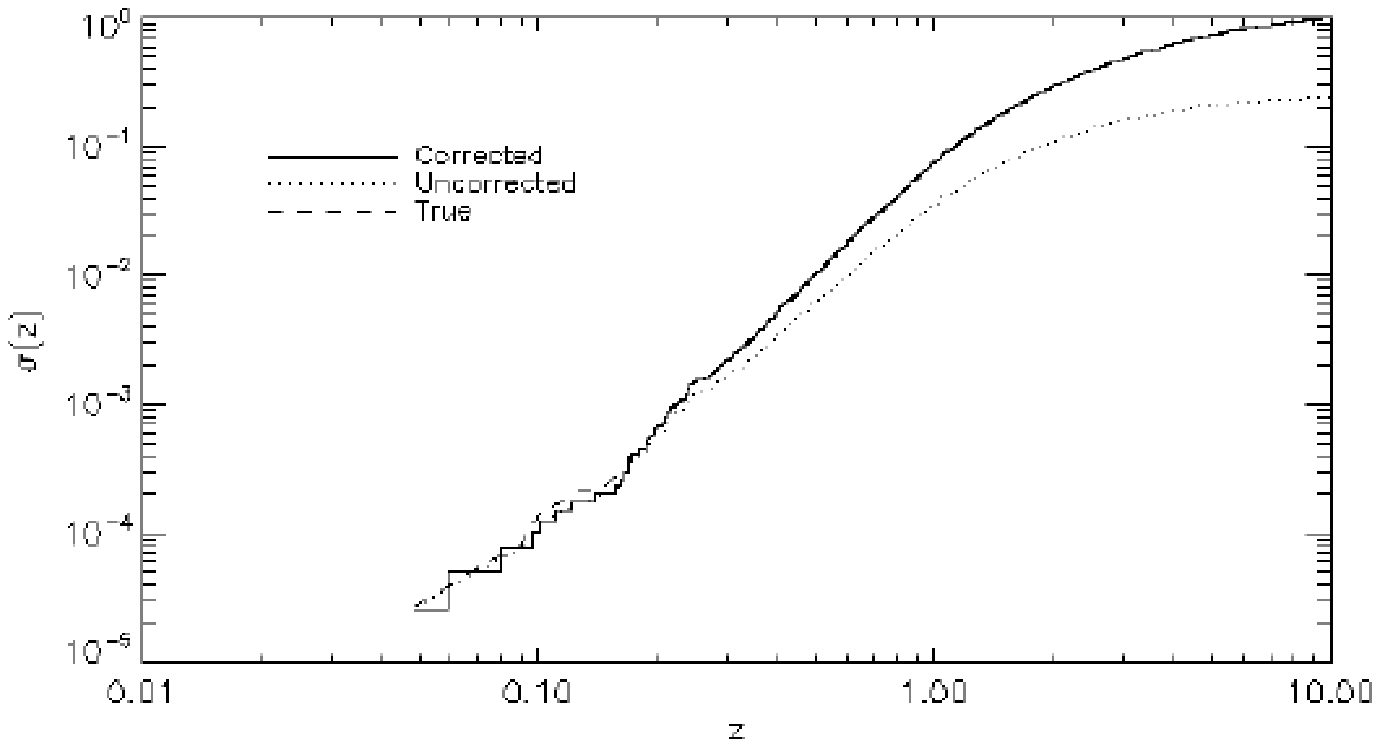}
\caption{Comparison of the true intrinsic cumulative distributions (dashed)of a
simulated sample (chosen to have characteristics similar to those of GRBs) with
that of the
flux limited ``observed" sample (dotted) and that histogram obtained by our
method that corrects
for this selection bias. As evident the method reproduces the input forms
extremely well, except at the extremities. {\bf Left:} LF $\phi(L)$.
{\bf Right:} density rate ${dot \sigma}(z)$.}
\label{sims}
\end{center}
\end{figure}

To demonstrate the accuracy of our methods we have simulated a sample of
sources with  redshift and luminosity distributions  similar to those of GRBs
with specified form for the LF and density rate evolution but with no
luminosity evolution. We
have selected a flux limited subsample and applied the
Efron-Petrosian methods to recover the intrinsic distributions.
We recover no evolution (obtain Kendell's $\tau = 0$ for $k\sim
0.1$), and as shown in Fig.~\ref{sims},
the calculated cumulative distributions $\phi(L)$ and $\sigma(z)$, agree very
well with the intrinsic distribution of the parent sample.

\section*{Appendix B: Non-parametric Differential Distribution}

As mentioned in \S 3 the Efron-Petrosian method gives a non-parametric and
point by point description of the cumulative distributions. Our usual procedure
is to fit such histograms to a smooth function and take its derivative to
obtain the differential distributions. This is because one cannot
differentiate a discontinuous histogram. Here we describe a point by point
estimation of the differential distributions (using luminosity as an example)
which can give us the histogram of the differential
distribution. 

The cumulative distribution up to and including each data point, say
$\phi(L_i)$ is given as 
\beq\label{LFSigma}
\phi(L_i)=\prod_{j=2}^i(1+1/N_j)
\eeq
where $N_i$ is the number of data points in the associated set of point $i$
(see Fig.\,\ref{LvsZ}). We can use two different estimator of the differential
distribution $\psi(L)=-d\phi(L)/dL$. 

We can use the estimator $\psi(L_i)=(\Phi(L_i)-\phi(L_{i-1})/(L_{i-1}-L_i)$ in
which  case it is easy to show that for $\delta_i\equiv L_{i-1}/L_i-1$ 
\beq\label{diff1}
L_i\psi(L_i)={\phi(L_i)\over \delta_iN_i}\times {1\over (1+1/N_i)}\simeq
{\phi(L_i)\over \delta_iN_i}(1-1/N_i+...), 
\eeq
where the last equality is for $N_i\gg 1$.

Another estimator is obtain using the fact that Eq.~(3) yields
$\delta
\ln\Phi(L_i)=\ln\phi(L_i)-\ln\phi(L_{i-1})=\ln(1+1/N_i)$, which then gives 
\beq\label{diff2}
L_i\psi(L_i)=\phi(L_i){d\ln\phi(L_i)\over d\ln L_i}= \phi(L_i){\ln(1+1/N_i)\over
\ln (1+\delta_i)}\simeq
{\phi(L_i)\over \delta_iN_i}(1-1/(2N_i)-\delta_i+ ...),
\eeq
which for $N_i\gg 1$ and $\delta_i\ll 1$ is equal to the first estimator.

We can similarly obtain the differential distribution $d{\dot \sigma}/dz$ and
${\dot\rho}(z)$.  

Unfortunately, for a sparse data set with sometimes large gaps, such as
that available for GRBs, not all $N_i$ are large (especially at the two ends
of the data, and the relevant quantities $\ln L-i$ or $\delta_i$ have large
dispersions. These fact introduce a large random noise so that we rely on
smoothing and differentiation of the cumulative distributions. The above
relation can be used for samples with dense and uniform coverage of the phase
space.

\bibliographystyle{apj}
\bibstyle{aa}


\def\refer { \par \noindent \hangindent=2pc \hangafter=1}
\baselineskip = 10 true pt

\section*{REFERENCES}
\refer Amati, L. et al. 2002, {\it A\&A},  390, 81
\refer Atteia J.-L. et al. 2004, AIP Conf. Proc. 727, p 37
\refer Band, D. L. \& Preece, R. D. 2005 \apj, 627, 319
\refer Bloom, J. S., Frail, D. A. \& Sari, R. 2001, {\it AJ}, 121, 2879
\refer Bouvier, A. 2010, PhD Thesis, Stanford University
\refer Butler, N. R.; Kocevski, D. \& Bloom, J. S. 2009, \apj, 694, 76
\refer	Butler, N. R.; Bloom, J. S.; Poznanski, D. 2010, \apj, 711, 495
\refer Dainotti, M. G., Petrosian, V. et al. 2013,  \apj, 774..157
\refer Dainotti, M.G., Del Vecchio, R., Nagataki, S. \& Capozziello, S., \apj,
2015, 800, 31
\refer Daigne, F. et al. 2006, {\it MNRAS}, 372, 1034
\refer Eddington {\it MNRAS}, 1915, 73, 359 and 1940,  100, 354
\refer Efron, B.  \& Petrosian, V.  1992, \apj, 399, 345 ({\bf EP})
\refer Efron, B.  \& Petrosian, V.  1999, J. Am. St. Assoc., 94, 824 
\refer Ghirlanda, G. et al. 2004, \apj, 613, L13
\refer Howell, E. J., Coward, D. M., Stratta, G., Gendre, B. \&
Zhou 2014 {\it MNRAS}, 444,  15
\refer  Hopkins, a. M.  \& Beacom, J. F. 2006, \apj, 651, 142
\refer Jakobsson , P. et al. 2006, {\it A\&A}, 447, 897
\refer Kistler, M. D.,Yuksel, H.,Beacom, J. F. \& Stanek,K. Z. 2008, \apjl, 673,
L119
\refer Kocevski, D. \& Liang, E.  2006, \apj, 642, 371
\refer Lamb, D.Q. \ea. 2004, {\it New Ast. Review}, 48, 459
\refer Le, T.  \& Dermer, C. D. 2007, \apj 661, 394
\refer Li,  L.-X. 2007, {\it MNRAS Letters}, 379, 55L
\refer Li,  L.-X. 2008, MNRAS, 388, 1487
\refer Lloyd, N. M.\& Petrosian, V. 1999, \apj, 511, 550
\refer Lloyd, N. M., Petrosian, V. \& Mallozzi, R.S. 2000, \apj, 534, 227
\refer Lloyd, N. M., Fryer, C. L. \& Ramirez-Ruiz, E. 2002, \apj,  574,554
\refer Lynden-Bell, D. 1971, {\it MNRAS}, 155, 95
\refer Malmquist, K. G. 1920, {\it Medd. Lund. Obs.}, Ser. 2, No. 22
\refer Maloney, A. \& Petrosian, V. 1999, \apj, 518, 32
\refer Nakar, E. \& Piran, T. 2004, {\it MNRAS}, 360, 73
\refer Natarajan, P. et al. 2005, {\it MNRAS}, 364, L8
\refer Neyman, J. \& Scott, E. L. 1959, {\it Handbuch Der Physik}, 
Ed. ~S. Flugge, Springer-Verlag, Berlin, 53, 416
\refer Norris, J. P., Marani, G.F. \& Bonnell, J.T. 2000, \apj, 534, 248
\refer Nysewander, M.  et al.  2009, \apj, 701, 824
\refer Peebles, P. J. E. {\it Principles of Physical Cosmology}, 1993,
Princeton Series in Physics, p330
\refer Petrosian, V. 1992, {\it Stat. Challenges in Modern Astro.}, p173
Eds. Feigelson \& Babu, (N.Y.  Springer-Verlag) 
\refer Petrosian, V. 1973, \apj 183, 359
\refer Petrosian, V. et al. 2009; arXiv-0909.5051
\refer Porciani, C. \& Madau, P. 2001, \apj, 548, 522
\refer Reichart, D. E. et al. 2001, 552, 57
\refer Robertson, B. E. \& Ellis, R.S. 2012, \apj, 744, 95
\refer Salvaterra, R. et al. 2009, {\it MNRAS}, 397, 602
\refer Salvaterra, R. et al. 2012, \apj, 794, 68
\refer Schmidt, M. 1968, \apj, 151, 393
\refer Singal, J., Petrosian, V. \& et al. 2011, \apj, 743, 104
\refer Singal, J., Petrosian, V. \& Ajello, M. 2012, \apj, 753, 45
\refer Singal, J., Petrosian, V. \& et al. 2013, \apj, 764, 43
\refer Trumpler, R. J. \& Weaver, H.F. 1953, {\it Statistical Astronomy}, Dover,
N.Y.
\refer Wanderman, D. \& Piran, T. 2010, {\it MNRAS}, 406, 1944
\refer Wei et al. 2014, {\it MNRAS} 3329, 439,
\refer Xiao, L. \& Schaefer, B. E. 2009, \apj, 707, 387
\refer Yonetoku, D., Murakami, T., Nakamura, T. R., Yamazaki, A. K. Inoue, \&
K. Ioka 2004 \apj, 609, 953

\end{document}